\documentclass[prd,twocolumn,reprint,preprintnumbers,nofootinbib]{revtex4-1}

\input{header}

\begin{document}
\title{Probing Light Gauge Bosons in Tau Neutrino Experiments}

\author{Felix Kling}
\email{felixk@slac.stanford.edu}
\affiliation{SLAC National Accelerator Laboratory, 2575 Sand Hill Road, Menlo Park, CA 94025, USA}

\begin{abstract}
The tau neutrino is probably the least studied particle in the SM, with only a handful of interaction events being identified so far. This can in part be attributed to their small production rate in the SM, which occurs mainly through $D_s$ meson decay. However, this also makes the tau neutrino flux measurement an interesting laboratory for additional new physics production modes. In this study, we investigate the possibility of tau neutrino production in the decay of light vector bosons. We consider four scenarios of anomaly-free $U(1)$ gauge groups corresponding to the $B\!-\!L$, $B\!-\!L_\mu\!-\!2L_\tau$, $B\!-\!L_e\!-\!2L_\tau$ and $B\!-\!3L_\tau$ numbers, analyze current constraints on their parameter spaces and explore the sensitivity of DONuT and as well as the future emulsion detector experiments FASER$\nu$, SND@LHC and SND@SHiP. We find that these experiments provide the leading direct constraints in parts of the parameter space, especially when the vector boson's mass is close to the mass of the $\omega$ meson. 
\end{abstract}

\maketitle

%%%%%%%%%%%%%%%%%%%%%%%%%%%%%%%%%%%%%
\section{Introduction}
%%%%%%%%%%%%%%%%%%%%%%%%%%%%%%%%%%%%%

The standard model (SM) consist of 17 particles, out of which the tau neutrino $\nu_\tau$ is the least experimentally constrained one. To detect the rare tau neutrino interactions, an intense neutrino beam with a sufficiently large beam energy to produce a tau lepton, $E_\nu \gtrsim 3.5~\gev$, is needed. Additionally, in order to identify a $\nu_\tau$ event, the neutrino detector needs to have sufficient spatial resolution to resolve the tau lepton decay topology. This is typically achieved using emulsion detectors~\cite{Nakamura:2006xs}, which have spatial resolutions down to $50~\nm$ and a correspondingly high number of detection channels of the order of $10^{14}/\cm^3$.

The world's dataset of directly observed tau neutrino interactions consists of 10 events observed at OPERA~\cite{Agafonova:2018auq} and 9 events observed at DONuT~\cite{Kodama:2007aa}. While tau neutrinos observed at OPERA are produced indirectly through $\nu_\mu \to \nu_\tau$ neutrino oscillations, tau neutrinos at DONuT are produced directly in inelastic collisions of the $800~\gev$ Tevatron proton beam with a tungsten target. 

More recently, additional emulsion-based experiments have been proposed which would be able to detect tau neutrino interaction events. Following the same approach as DONuT, the scattering and neutrino detector of the SHiP experiment (SND@SHiP) would be located at a SPS beam dump facility and could detect about 11,000 tau neutrino interactions within 5 years of operation~\cite{Anelli:2015pba, Ahdida:2654870, Ahdida:2704147}. The recently approved FASER$\nu$ detector will be placed about $480~\m$ downstream from the ATLAS interaction point, where it utilizes the LHC's intense neutrino beam, and is expected to detect about 11 $\nu_\tau$ interactions by 2023~\cite{Abreu:2019yak, Abreu:2020ddv}. Following the same general idea, but placed on the other side of the ATLAS interaction point, the proposed SND@LHC detector could also detect a similar number of events in the same time~\cite{Ahdida:2020evc}.

In the SM, tau neutrinos are mainly produced in the decay of $D_s$ mesons, leading to a small flux compared to other neutrino flavors. This small SM production rate makes the tau neutrino flux measurement an interesting laboratory for additional beyond the SM (BSM) production modes.  

One example of such new physics are light vector bosons $V$ associated with additional gauge groups. These can be abundantly produced in meson decays, such as $\pi^0 \to V\gamma$, and then decay into neutrinos, $V \to \nu\nu$. Most importantly, light vector bosons often decay roughly equally into all three neutrino flavors, leading to a sizable tau neutrino flux. This contribution could, in principle, be comparable to or larger than the SM $\nu_\tau$ flux and, hence, allows us to probe such models in tau neutrino experiments. 

The rest of this paper is organized as follows. In \autoref{sec:model} we will discuss light vector boson models. In \autoref{sec:constraints} we will discuss existing constraints on these models before analyzing the sensitivity of tau neutrino measurements in \autoref{sec:sensitivity}. We conclude in \autoref{sec:summary}. 

%%%%%%%%%%%%%%%%%%%%%%%%%%%%%%%%%%%%%
\section{Additional Vector Bosons}
\label{sec:model}
%%%%%%%%%%%%%%%%%%%%%%%%%%%%%%%%%%%%%

The Lagrangian of the SM contains four linearly independent global symmetries corresponding to the baryon number, $U(1)_B$, and the individual lepton family numbers, $U(1)_{L_e}$, $U(1)_{L_\mu}$ and $U(1)_{L_\tau}$. One of the simplest ways to extend the SM is to gauge anomaly-free combinations of these global symmetries. This includes the difference between lepton family numbers $L_i\!-\!L_j$, with $i,j=e,\mu,\tau$, and the difference between baryon and lepton number $B\!-\!L$, where the latter requires the addition of three right-handed neutrinos to the SM to guarantee anomaly cancellation. 

Additionally, any linear combination of the anomaly-free groups $U(1)_{L_i-L_j}$ and $U(1)_{B-L}$ will also be anomaly-free. This leads to two general classes of anomaly-free groups corresponding to $x_e L_e\!+\! x_\mu L_\mu \!-\! (x_e\!+\!x_\mu) L_\tau$ and $B\!+\! x_e L_e \!+\! x_\mu L_\mu \!-\! (3 \!+\! x_e \!+\! x_\mu) L_\tau $, where $x_e$ and $x_\mu$ are real numbers~\cite{Araki:2012ip, Asai:2019ciz}.

In all of these cases, a new vector boson $V$ is introduced. It couples to the standard model fermions proportionally to the gauge group's coupling constant, $g$, and the fermion charges under the $U(1)$ symmetry, $q_f$. In \autoref{tab:groups}, we summarize the fermion charges $q_f$ for the general case, as well as for the gauge groups considered in this study. The Lagrangian for this model can then be written as
\be
 \mathcal{L} = \mathcal{L}_{SM} - \frac{1}{2} m_{V}^2 V_\mu V^\mu -  g \sum q_{f} V^\mu \bar{f_i} \gamma_\mu f_i  \ ,
\ee
where $\mathcal{L}_{SM}$ is the Lagrangian of the SM and $m_V$ is the mass of the new vector boson. Once the gauge group is fixed, the parameter space of the model is spanned by the gauge boson's mass $m_V$ and the coupling $g$. 

In this study, we are mainly interested in the scenarios that allow for efficient gauge boson production in hadron collisions and subsequent decay into tau neutrinos. We therefore choose to focus on models with couplings to quarks and taus and consider the following four anomaly-free groups: $B\!-\!L$, $B\!-\!L_e\!-\!2L_\tau$, $B\!-\!L_\mu\!-\!2L_\tau$ and $B\!-\!3L_\tau$. The corresponding fermion charges under these groups are also shown in \autoref{tab:groups}. In order to provide a sizeble production rate of these new states, we will focus on light particles with masses in the range $1~\mev < m_V < 10~\gev$. 

In principle, additional couplings could be induced though loop effects. In particular, fields charged under both the new $U(1)$ and the $U(1)_Y$ hypercharge groups will induce a kinetic mixing between the two groups, $\mathcal{L} \supset \epsilon \,  B_{\mu\nu} V^{\mu\nu}$, where $B_{\mu\nu}$ and $V_{\mu\nu}$ denote the field strength of the hypercharge boson $B$ and the new gauge boson $V$, respectively. However, as discussed in Ref.~\cite{Bauer:2018onh}, the kinetic mixing parameter $\epsilon$ cannot be determined unambiguously in the models considered. We will therefore neglect the kinetic mixing, but note that its presence could lead to additional constraints. 

Finally, we want to note that Ref.~\cite{ Bahraminasr:2020ssz} considers a similar scenario with a new gauge boson that couples only to neutrinos. However, this particular model is not invariant under $SU(2)_L$, and suffers from a low production rate due to a lack of direct couplings to hadrons. 

%------------------------
\begin{table}[t!]
\centering
\begin{tabular}{c|c|c|c|c}
  \hline
  \hline
  Gauge Group	 		
  & $q_{q}$     & $q_{e}$  		& $q_{\mu}$		& $q_{\tau}$    \\
  \hline
  \hline
  $x_e L_e\!+\!x_\mu L_\mu\!-\!(x_e\!+\!x_\mu) L_\tau$
  & 0			& $x_e$ 		& $x_\mu$ 		& $\!-\!x_e\!-\!x_\mu$ 
  \\ 
  $B\!+\!x_e L_e\!+\!x_\mu L_\mu\!-\!(3\!+\!x_e\!+\!x_\mu) L_\tau$
  & 1/3			& $x_e $		& $x_\mu$ 		& $\!-3\!-\!x_e\!-\!x_\mu$ 
  \\
\hline	
  $B\!-\!L$			
  & 1/3			& $-1$			& $-1$			& $-1$		    \\
  $B\!-\!L_\mu\!-\!2L_\tau$ 
  & 1/3			& 0				& $-1$			& $-2$		    \\
  $B\!-\!L_e\!-\!2L_\tau$	
  & 1/3			& $-1$			& 0				& $-2$		    \\
  $B\!-\!3L_\tau$		
  & 1/3			& 0				& 0				& $-3$		    \\
 \hline
 \hline
\end{tabular}
\caption{Types of anomaly-free gauge groups and corresponding fermion charges $q_f$.}
\label{tab:groups}
\end{table}
%------------------------

%%%%%%%%%%%%%%%%%%%%%%%%
\section{Existing constraints}
\label{sec:constraints}
%%%%%%%%%%%%%%%%%%%%%%%%

Light vector bosons models have a rich phenomenology, whose details depend on the particle's mass and coupling as well as the underlying group structure. In the following, we will discuss both direct searches looking for $s$-channel production and indirect searches using $t$-channel exchange of the vector boson.  

\subsection{Direct Dark Photon Searches}

Many experiments have performed direct searches for light vector bosons, in which an on-shell vector boson is produced. Their results are typically presented in the context of searches for a dark photon, which kinetically mixes with the SM photon, leading to couplings of the dark photon to SM fermions proportional to their electric charge. For most constraints, we use the \texttt{DarkCast} tool~\cite{Ilten:2018crw} to recast these dark photon limits and obtain the bounds for the models considered in this study. The resulting limits are shown in \autoref{fig:results} as dark gray shaded regions. 

\begin{description}

\item[Prompt Decays]

Searches for visibly decaying dark photons have been performed at a large variety of fixed target and collider experiments with both electron and hadron beams. If the coupling $g$ is large, the vector boson $V$ decays promptly in the detector. The resulting resonant signal can be identified over the typically continuous background by performing a bump hunt over the invariant mass spectrum. 

The most important bounds have been obtained by the dark photon search for $ee \to A' \gamma$ with $A' \to ee,\mu\mu$ at BaBar~\cite{Lees:2014xha}; the dark photon search for $A' \to \mu\mu$ at LHCb~\cite{Aaij:2019bvg}; and the search for a dark photon in the decay $Z \to A' \mu \mu \to 4 \mu$ at CMS~\cite{Sirunyan:2018nnz} as discussed in Ref.~\cite{Chun:2018ibr}.

\item[Displaced Decays]

In contrast, if the coupling $g$ is small, the vector boson's lifetime becomes large, $\tau_V \sim g^{-2} m_V^{-1}$, and $V$ will decay a macroscopic distance away from where it is produced. Many fixed target and beam dump experiments have searched for such displaced decays occurring in a detector placed downstream from the collision point. Because of additional shielding in front of the detector, these searches can be performed in a low-background environment which allows them to probe the small coupling regime with small associated event rates. 

The most sensitive constraints have been obtained by searches for dark photon decays $A' \to ee$ using the proton beam dump  experiment NuCal~\cite{Blumlein:1990ay, Blumlein:1991xh} and the electron beam dump experiment Orsay~\cite{Davier:1989wz}. 

\item[Invisible Decays]

The gauge groups considered in this study are designed to have a large branching fraction into neutrinos. This decay will lead to missing energy signatures, which have been probed by various experiments searching for dark photon decays into dark matter. 

The most sensitive constraints have been obtained by the search for dark photon production $ee \to A' \gamma$ at BaBar~\cite{Lees:2017lec}; the search for dark photon production $eN \to eN A'$ at NA64~\cite{NA64:2019imj}; the search for the decay $\pi^0 \to \gamma A'$ at NA62~\cite{CortinaGil:2019nuo} and LESB~\cite{Atiya:1992sm}; the search for the decay $\pi^0,\eta,\eta' \to \gamma A'$ at Crystal Barrel~\cite{Amsler:1994gt}; the search for the decay $K^+ \to \pi^+ A'$ at E949~\cite{Artamonov:2009sz} as discussed in Ref.~\cite{Pospelov:2008zw, Batell:2014yra}; and the monojet search $pp \to A'+jet$ at CDF~\cite{Aaltonen:2012jb} as discussed in Ref.~\cite{Shoemaker:2011vi}. 

\end{description}

\subsection{Indirect Constraints}

In addition to direct searches, many indirect constraints arise from both scattering experiments probing the exchange of the light vector boson as well as precision measurements sensitive to induced radiative corrections. 

Although indirect searches provide a powerful tool to search for new physics, our interpretation typically relies on the additional underlying assumptions that no further new physics is present or interferes with the considered light vector boson contribution. These constraints should therefore be considered as model dependent, and it should be noted that they could be relaxed in the presence of additional new physics. In the following, we summarize the most important constraints and recast them for our four models. The resulting limits are shown in \autoref{fig:results} as light gray shaded regions enclosed by dashed lines.

\begin{description}

\item [Neutrino Cross Sections]

Light vector bosons with couplings to neutrinos can modify neutrino scattering cross sections, which can therefore be used to constrain such models. The most sensitive constraints are imposed by the measurement of the neutrino trident production rate $\nu_\mu N \to \nu_\mu \mu \mu N$ for models with $q_\mu\neq 0$ by CCFR~\cite{Mishra:1991bv} as discussed in Ref.~\cite{Altmannshofer:2014pba}; the measurement of the cross section for $\nu_\mu e \to \nu_\mu e$ scattering for models with $q_e,q_\mu \neq 0$ by CHARM-II~\cite{Vilain:1993kd} as discussed in Ref.~\cite{Bilmis:2015lja}; and the measurement of the cross section for coherent neutrino scattering on a CsI target $\nu_\mu N \to \nu_\mu N$ for models with $q_\mu \neq 0$ by  COHERENT~\cite{Akimov:2017ade} as discussed in Ref.~\cite{Kosmas:2017tsq}. 

\item [Muon Anomalous Magnetic Moment]

The anomalous magnetic moment of the muon, $a_\mu$, is one of the most precisely measured quantities in particle physics. Interestingly, the experimentally measured value $a_\mu^{exp}$ differs from its SM prediction $a_\mu^{SM}$ by an amount~\cite{Bennett:2006fi, Tanabashi:2018oca}
\be
 \Delta a_\mu = a_\mu^{exp} - a_\mu^{SM} = (26.1 \pm 7.8) \times 10^{-10} \ . 
\ee 
While this measurement puts a constraint on models of new physics, it also motivates the existence of light new particles to explain the anomaly. In \autoref{fig:results}, we show the $2\sigma$ region of parameter space accommodating the anomaly, $10.4 \times 10^{-10}< \Delta a_{\mu} < 41.8 \times 10^{-10}$, as green shaded bands. Large $\Delta a_\mu> 65.1 \times 10^{-10}$ are excluded at the $5\sigma$ level. 
	
\item [LEP Z-pole Measurements] 

$Z$-pole measurements at LEP have determined the leptonic decay widths of the $Z$-boson with high precision~\cite{Tanabashi:2018oca}. In particular, these measurements constrain any BSM contribution to the decay width into tau leptons,
\be
\Delta\Gamma_{Z \to \tau\tau}^{\text{BSM}} / 
\Gamma_{Z \to \tau\tau} < 0.0046
\ee
at 95\% C.L., which would be modified in the presence of a new vector boson with couplings to taus~\cite{Ma:1998dp}. 

\item [Neutron Scattering Measurements] 

The existence of a new vector boson can also be probed by low-energy nuclear scattering experiments. In particular, the measurement of the differential cross section for neutron-lead scattering with a neutron beam energy between 1 and 26 keV~\cite{Barbieri:1975xy} provides a constraint on the vector boson parameter space $q_{n,p} \cdot g < (m_V / 206~\mev)^2$~\cite{Barger:2010aj}, where $q_{n,p}=1$ are the neutron and proton charges under the groups considered in this paper.

\item [Non-Standard Interactions] 

A series of neutrino experiments have measured neutrino oscillations both in vacuum and in the matter background of the Sun and Earth. A combination of these results allows to put constraints on non-standard interactions (NSI) between neutrinos and matter, which are traditionally parameterized through terms $\propto \epsilon_{ii}^f (\bar\nu_i \gamma_\mu \nu_i )(\bar f \gamma^\mu f)$. A global fit~\cite{Esteban:2018ppq} to neutrino oscillations has constrained the difference between the NSI of the muon and tau neutrino with nuclear matter
\be
-0.008<\epsilon^{n+p}_{\tau\tau} - \epsilon^{n+p}_{\mu\mu} <0.18 
\ee
which provides the strongest constraint on the light vector boson models with baryon couplings considered in this paper. Following Ref.~\cite{Heeck:2018nzc}, we can estimate the vector boson's contribution to NSI as 
\be
\epsilon^{n+p}_{\tau\tau} - \epsilon^{n+p}_{\mu\mu} =  -\frac{g^2}{2 \sqrt{2} G_F m_V^2} (3 + x_e + 2 x_\mu )
\ee
This constraint is most relevant for the otherwise poorly constrained $B\!-\!3L_\tau$ and $B\!-\!L_\mu\!-\!2L_\tau$ models and is shown with a dotted contour in \autoref{fig:results}.

\end{description}

In addition to these existing constraints, a series of future searches and experiments have been proposed to probe the parameter space of light vector bosons. These experiments and their estimated reach for dark photons, $B\!-\!L$ and $L_i\!-\!L_j$ gauge bosons are discussed in detail in Refs.~\cite{Bauer:2018onh, Foldenauer:2019dai}. 

%%%%%%%%%%%%%%%%%%%%%%%%%%%%%%%%%%%%%
\section{Tau Neutrino Measurements}
\label{sec:sensitivity}
%%%%%%%%%%%%%%%%%%%%%%%%%%%%%%%%%%%%%

%------------------------
\renewcommand{\arraystretch}{1.2}
\setlength{\tabcolsep}{3.4pt}

\begin{table*}[t!]
\centering
\begin{tabular}{c|c|c|c|c|c|c||c|c||c|c|c}
  \hline
  \hline
  \multicolumn{7}{c||}{Experimental Setup} & 
  \multicolumn{2}{c||}{SM} & 
  \multicolumn{3}{c}{$B\!-\!3L_\tau$} \\
  \hline
  Experiment	
  & Status    
  & $\mathcal{L}$/$N_{\text{POT}}$ 
  & $m_{\text{det}}$
  & $A_\text{det}$ 
  & $\epsilon_\text{det}$    
  & Ref. 
  & $N_{\text{event}}$ 
  & $\langle E_\nu \rangle$
  & $N_{\text{event}}$ 
  & $\langle E_\nu \rangle$ 
  & $N_{2\sigma}$
  \\
  \hline
  DONuT         
  & completed 
  & $3\cdot10^{17}$   
  & $0.26~\ton$
  & $50 \times 50~\cm^2$
  & 0.2   
  & \cite{Kodama:2007aa} 
  & $10 \pm 4.6$ 
  & $112~\gev$
  & 12
  & $\phantom{0}84~\gev$
  & $9.1$ 
  \\
  FASER$\nu$    
  & approved  
  & $150~\ifb$      
  & $1.2~\ton$
  & $25 \times 25~\cm^2$
  & 0.52 
  & \cite{Abreu:2019yak} 
  & $11.6 \pm 5.1$ 
  & $965~\gev$ 
  & 96
  & $928~\gev$
  & $10$ 
  \\
  SND@LHC       
  & proposed  
  & $150~\ifb$    
  & $0.85~\ton$
  & $40 \times 40~\cm^2$
  & 0.5   
  & \cite{Ahdida:2020evc} 
  & $4.3 \pm 2.5$
  & $720~\gev$
  & 3.5
  & $382~\gev$
  & $5 $ 
  \\
  SND@SHiP      
  & proposed  
  & $2\cdot10^{20}$   
  & $8~\ton$
  & $80 \times 80~\cm^2$
  & 0.22  
  & \cite{Ahdida:2654870} 
  & ($10.9 \pm 3.6)\cdot 10^3$
  & $\phantom{0}52~\gev$
  & $2\cdot 10^4$
  & $\phantom{0}54~\gev$
  & $7200 $ 
  \\
  \hline
  \hline
\end{tabular}
\caption{Comparison of the experiments and their expected event numbers. The first block summarizes the experimental setup, including the status of the experiment, the assumed luminosity at the LHC $\mathcal{L}$ or the number of protons on target $N_{\text{POT}}$ at proton beam dump experiments, the mass of the detector $m_{\text{det}}$, the detector's cross sectional area $A_\text{det}$, the efficiency to detect tau neutrinos $\epsilon_\text{det}$, and the reference used. The second block shows the expected number of observable tau neutrino events from $D_s$ meson decay and their average energy. The last block shows the expected number of observable tau neutrino events from the decay of a $B\!-\!3L_\tau$ gauge boson with mass $m_V=10~\mev$ and coupling $g=10^{-3}$, the average neutrino energy and the number of events to exclude a parameter point in this model at $2\sigma$, $N_{2\sigma}$.
}
\label{tab:experiments}
\end{table*}
%------------------------

%----------------------------------
\subsection{Experimental Setup}
%----------------------------------

In this study we consider four experiments which are able to identify tau neutrino interactions and, hence, constrain BSM production modes: DONuT, FASER$\nu$, SND@SHiP and SND@LHC. Below, we briefly review each experiment and summarize their characteristics relevant for this study. 

DONuT~\cite{Kodama:2000mp, Kodama:2007aa} was an experiment at Fermilab designed to detect tau neutrinos for the first time. It utilized the $800~\gev$ proton beam of the Tevatron accelerator, which was directed into a fixed tungsten target. A detector consisting of $260~\kg$ of nuclear emulsion was placed $36~\m$ behind the interaction point and centered on the beam axis. By the end of operation, 9 $\nu_\tau$ events were identified, agreeing with the prediction of 10 events in the SM.

FASER is a new experiment at the LHC, which is located in the very forward direction. While its main focus is the search for light long-lived particles at the LHC~\cite{Feng:2017vli, Feng:2017uoz, Feng:2018noy, Kling:2018wct, Berlin:2018jbm, Ariga:2018zuc, Ariga:2018uku, Ariga:2018pin, Ariga:2019ufm}, the FASER experiment also contains an emulsion detector, FASER$\nu$, which has been designed to detect neutrinos at the LHC for the first time and consists of emulsion films interleaved with tungsten plates. The FASER experiment is located about $480~\m$ downstream from the ATLAS interaction point in the previously unused side tunnel TI12. At this location, a trench has been dug, which allows one to center both the FASER main detector and the FASER$\nu$ neutrino detector on the beam collision axis, covering the pseudorapidity range $\eta \gtrsim 9$. The FASER$\nu$ detector will collect data during run 3 of the LHC, from 2021 to 2024, which has a nominal luminosity of $150~\ifb$ and nominal center of mass energy of $14~\tev$. 

SHiP is a proposed high-intensity beam dump experiment using CERN's $400~\gev$ SPS beam and expected to collect $N_\text{POT}=2 \cdot 10^{20}$ protons on target. Its primary purpose is the search for long-lived particles~\cite{Alekhin:2015byh} using its hidden sector spectrometer. In addition, the SHiP proposal contains the scattering and neutrino detector, here called SND@SHiP, which would be able to record about 10,000 $\nu_\tau$ interactions~\cite{Anelli:2015pba, Ahdida:2654870, Ahdida:2704147}. The SND@SHiP detector is located about $46~\m$ behind the interaction point and is centered on the beam axis. 

More recently, the SHiP Collaboration proposed a similar detector design to be placed in the forward direction at the LHC. The SND@LHC~\cite{Ahdida:2020evc} detector would be placed in the tunnel TI18, which is also $480~\m$ away from the ATLAS interaction point, but on its other side. Notably, the center of the detector would be displaced from the beam collision axis by $28~\cm$ in the horizontal direction and $34~\cm$ in the vertical direction, providing a pseudorapidity coverage $7.2< \eta <8.7$ complementary to the FASER$\nu$ detector. The detector would also operate during run 3 of the LHC. 
\medskip 

In the left block of \autoref{tab:experiments} we summarize the experimental setup for each detector, including their assumed luminosity $\mathcal{L}$ or number of protons on target $N_{\text{POT}}$, detector mass $m_\text{det}$, cross sectional area $A_\text{det}$ and $\nu_\tau$ identification efficiency $\epsilon_\text{det}$. More information can be found in the listed references.

%------------------------------------------------------------
\begin{figure*}[tb]
\centering
\includegraphics[width=0.49\textwidth]{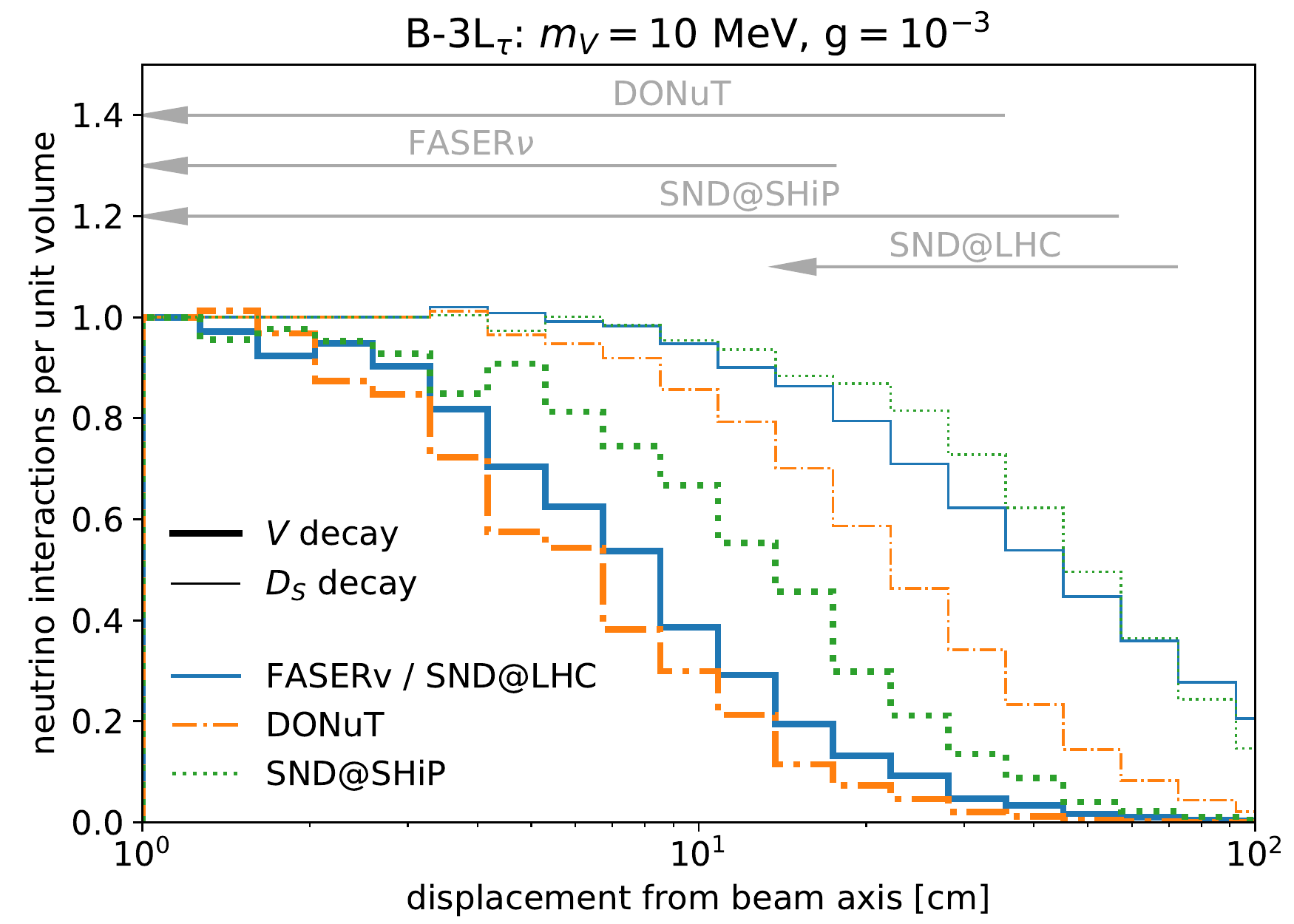}
\includegraphics[width=0.49\textwidth]{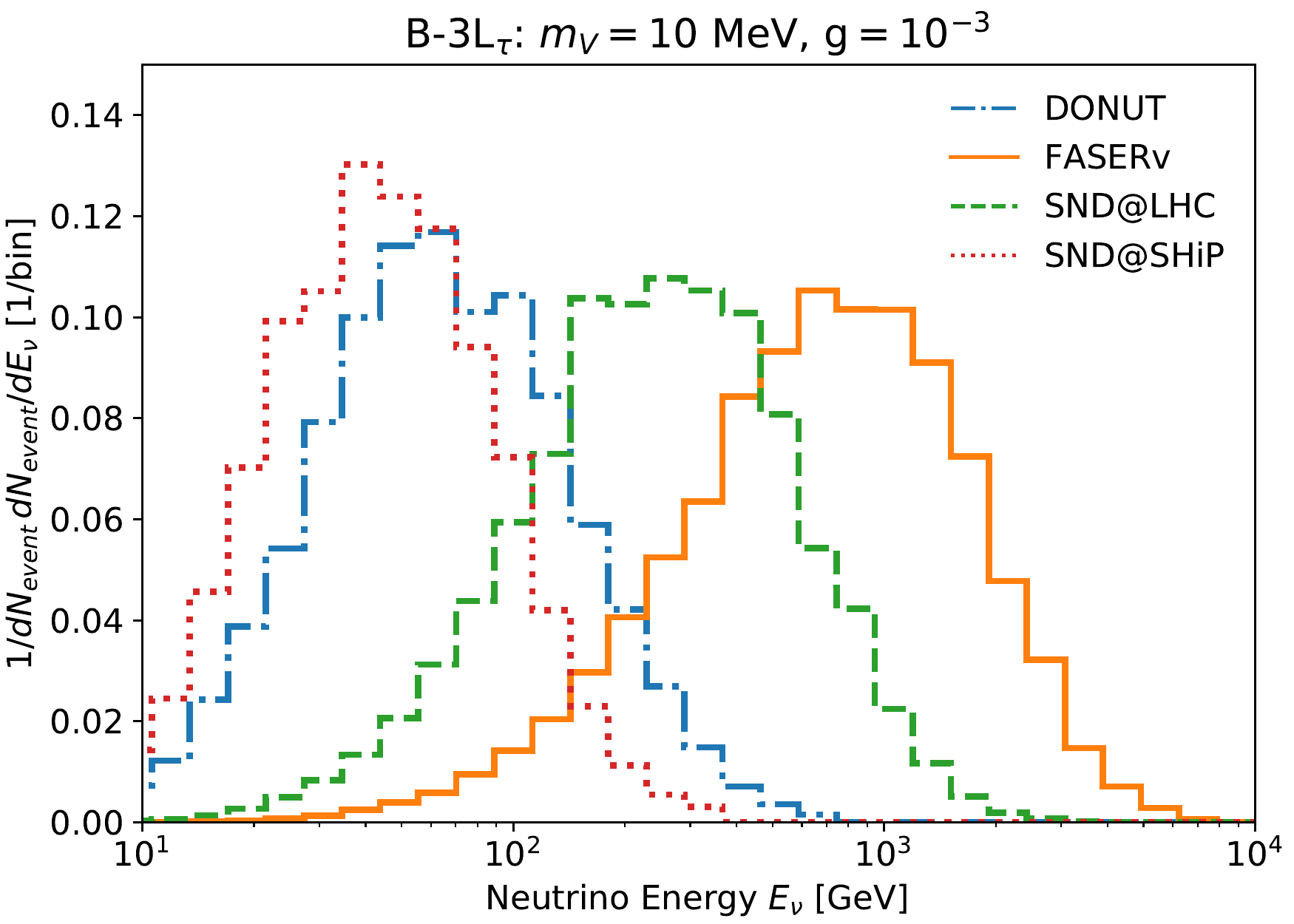}
\caption{Kinematic distributions of tau neutrinos produced in the decay of a $U(1)_{B\!-\!3L_\tau}$ vector boson with mass $m_V=10~\mev$ and coupling $g=10^{-3}$ at the considered experiments. The left panel shows the possible rate of neutrino interactions per unit volume, normalized to its value at the beam axis, as a function of the displacement from the beam axis. For comparison, we also show the radial distribution for tau neutrinos from $D_s$ decay. The gray arrows indicate the radial coverage of the considered experiments. The right panel shows the normalized energy distribution for $\nu_\tau$'s from $V$ decay interacting in the detector. 
}
\label{fig:distributions}
\end{figure*}
%------------------------------------------------------------

In the SM, tau neutrinos are mainly produced through the decay $D_s \to \tau \nu_\tau$ and the subsequent decay $\tau \to \nu_\tau + X$. In the center block of \autoref{tab:experiments}, we show the corresponding number of $\nu_\tau$ events expected in the SM for each detector\footnote{To allow for a fair comparison with FASER$\nu$, we have re-evaluated the event rate for SND@LHC using the more modern event generators \texttt{Sibyll~2.3c}~\cite{Engel:2015dxa, Fedynitch:2018cbl} and \texttt{Pythia~8}~\cite{Sjostrand:2014zea} with the \texttt{Monash} tune~\cite{Skands:2014pea} and \texttt{A2} tune~\cite{ATL-PHYS-PUB-2012-003}, as outlined in Ref.~\cite{Abreu:2019yak}. Compared to Ref.~\cite{Ahdida:2020evc}, which uses the pre-LHC event generator \texttt{DPMJET-III}~\cite{Roesler:2000he}, the expected number of events is reduced by roughly a factor two.}. Following the DONuT analysis~\cite{Kodama:2007aa}, we assume a 33\% systematic uncertainty for the SM tau neutrino flux in all experiments, which is added in quadrature to the statistical uncertainties. Dedicated theoretical efforts~\cite{Bai:2018xum, Bai:2020ukz} or direct measurements of $D_s$-meson production~\cite{Aoki:2019jry, Aaij:2015bpa} will play an important role in further reducing these uncertainties in the future. Also shown is the average energy of the tau neutrinos interacting with the detector. 

%----------------------------------
\subsection{Simulation}
%----------------------------------

We perform a dedicated Monte Carlo study to estimate the additional contribution to the neutrino flux from light vector boson decay. 

If the vector boson is light, it can be produced in the decay of light mesons. In particular, we take into account the decays $\pi^0, \eta,\eta' \to V \gamma$ and $\omega,\phi \to V \eta$. We generate the meson spectra using \texttt{EPOS-LHC}~\cite{Pierog:2013ria} as implemented in the simulation package \texttt{CRMC}~\cite{CRMC} and subsequently decay the mesons using the branching fractions obtained in Ref.~\cite{Tulin:2014tya}.

A heavier vector boson can be produced through bremsstrahlung $pp \to ppV$, which we model using the Fermi-Weizs{\"a}cker-Williams (FWW) approximation, following the procedure outlined in Ref.~\cite{Feng:2017uoz}. Note that the vector bosons with equal couplings to all quark flavors considered in this paper do not mix with the $\rho$~meson~\cite{Tulin:2014tya, Ilten:2018crw}. Therefore only the $\omega$-meson contribution is taken into account in the proton form factor used in the FWW approximation, leading to an enhanced production at $m_V \approx m_\omega$. For masses $m_V>1.7~\gev$, we additionally include vector boson production in hard scattering $qq \to V$, which we simulate with \texttt{Pythia~8}~\cite{Sjostrand:2014zea, Sjostrand:2006za}. 

In the next step, we decay the vector boson into tau neutrinos using the branching fractions provided by \texttt{DarkCast}~\cite{Ilten:2018crw}. Note that in the relevant region of parameter space, the vector boson's lifetime is short such that it will always decay promptly. The resulting distribution corresponds to the differential tau neutrino flux $d^2 N_{\nu}/dE_\nu\,d\theta_\nu$, where $E_\nu$ and $\theta_\nu$ are the neutrino energy and angle with respect to the beam axis, respectively.

The probability of the neutrinos interacting with the detector can be written as 
\be
 P_{\text{int}}(E_\nu,\theta_\nu) = \frac{\sigma_{\nu N}(E_\nu)}{
A_\text{det}} \frac{m_{\text{det}}}{m_N} \times \mathcal{A}(\theta_\nu)
\ee
where $\sigma_{\nu N} (E_\nu)$ is the energy-dependent neutrino interaction cross section with the target material~\cite{Abreu:2019yak}, $A_\text{det}$ is the detector's cross sectional area, $m_\text{det}$ is the detector's mass, $m_N$ is the mass of a target nucleus and $\mathcal{A}(\theta_\nu)$ corresponds to the angular acceptance of the detector. Finally, we obtain the $\nu_\tau$ event rate $N_\text{event}$ by convoluting the tau neutrino flux with the interaction probability and the detector's efficiency to identify tau neutrinos $\epsilon_\text{det}$, 
\be
\!N_\text{event} \!=\! \int\! \frac{d^2N(E_\nu,\theta_\nu)}{d E_\nu \ d \theta_\nu} 
\cdot  P_{\text{int}}(E_\nu,\theta_\nu) 
\cdot \epsilon_\text{det} \, dE_\nu \, d\theta_\nu \, . 
\ee
The detector's efficiency $\epsilon_\text{det}$, mass $m_\text{det}$ and area $A_\text{det}$ are given in \autoref{tab:experiments}.

%------------------------------------------------------------
\begin{figure*}[t]
\centering
\includegraphics[width=0.49\textwidth]{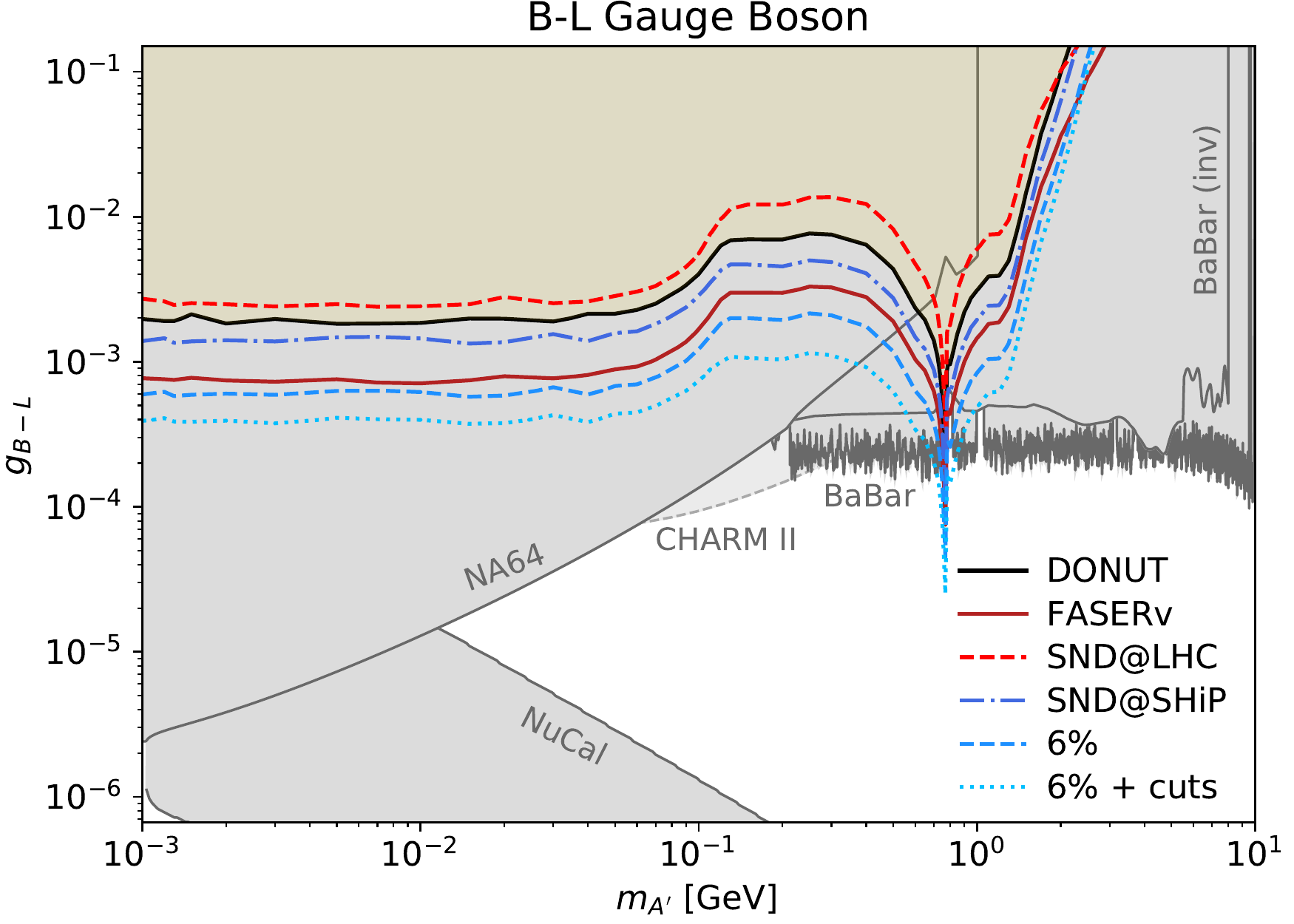}
\includegraphics[width=0.49\textwidth]{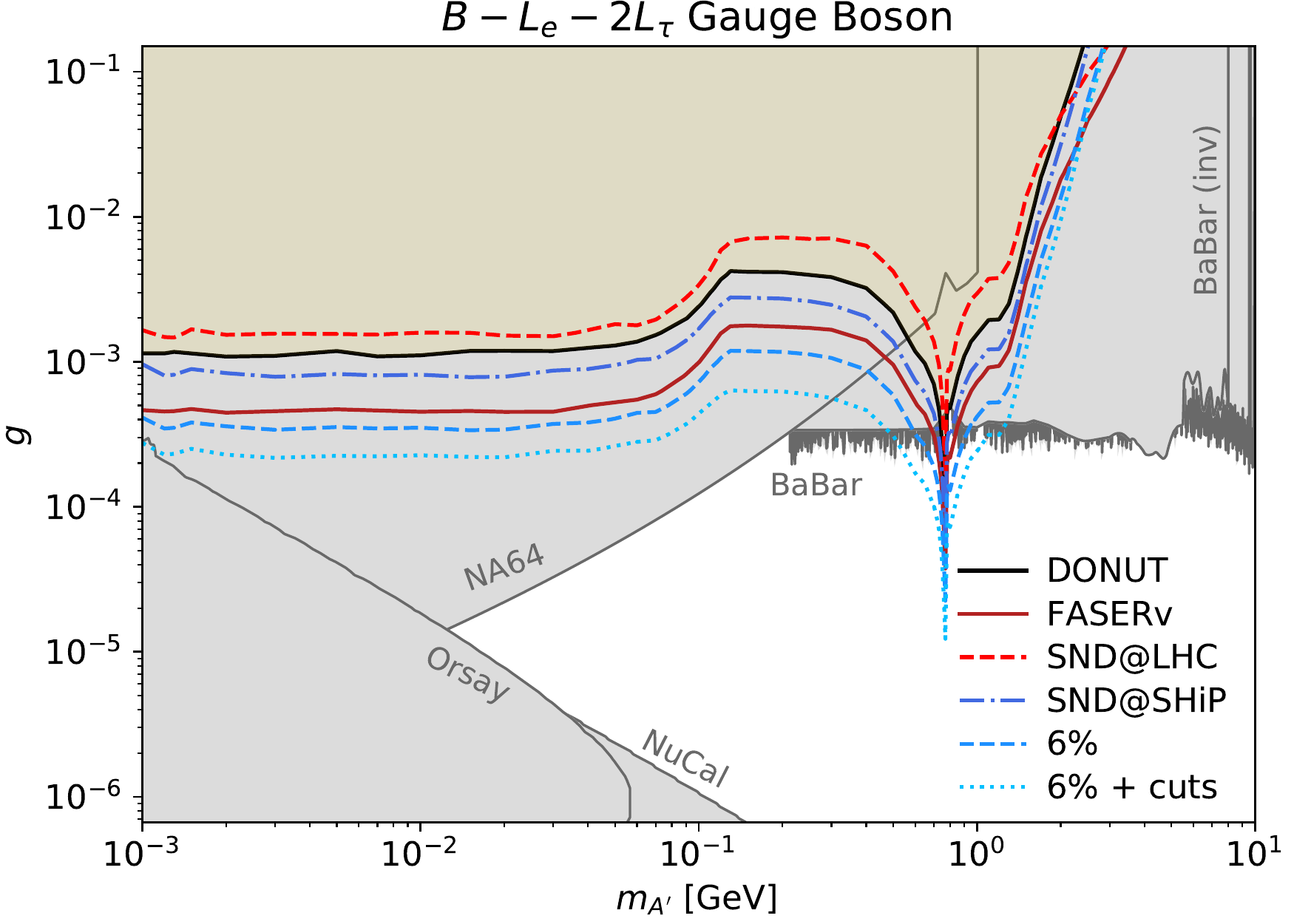}\\
\vspace{0.3cm}
\includegraphics[width=0.49\textwidth]{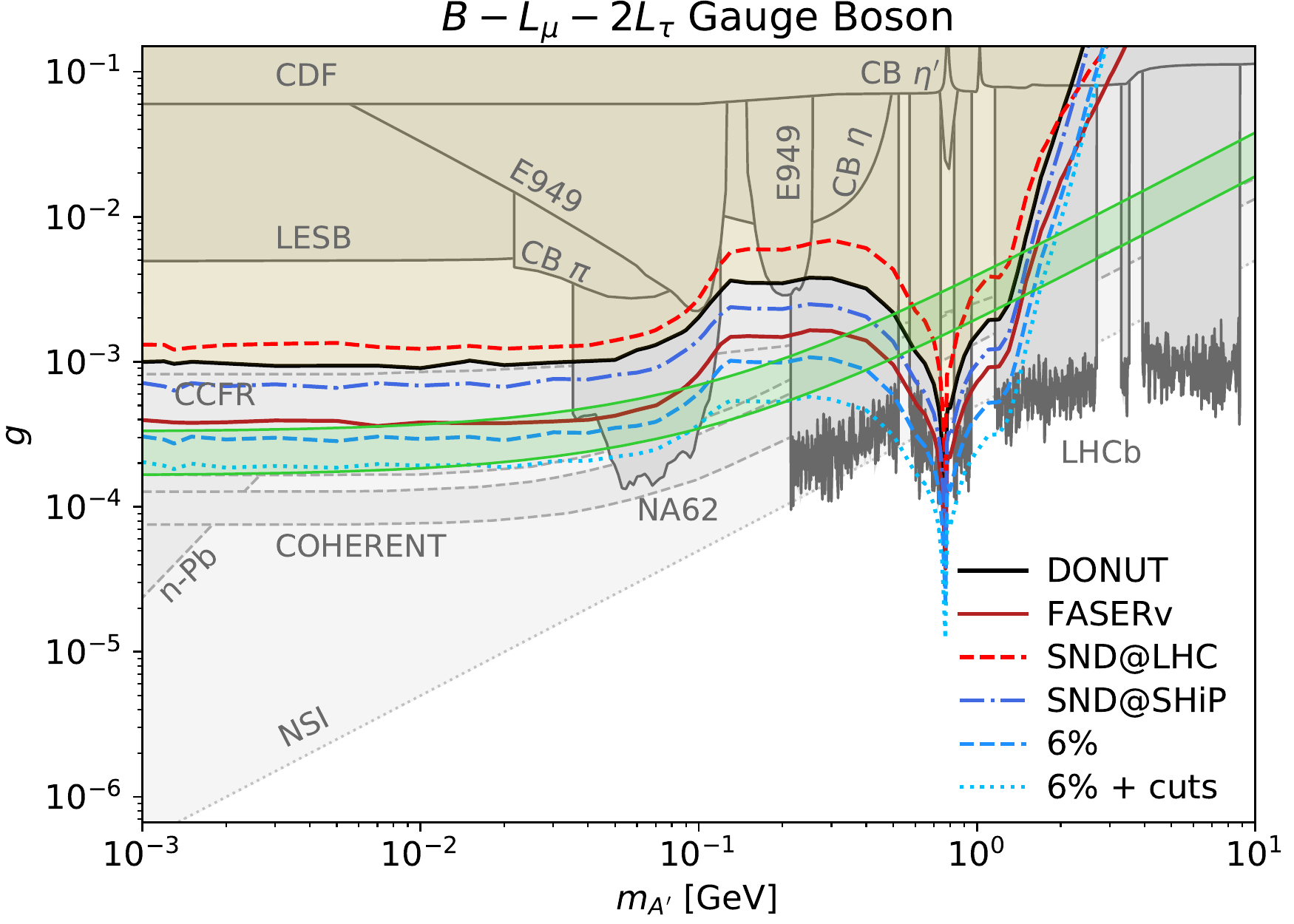}
\includegraphics[width=0.49\textwidth]{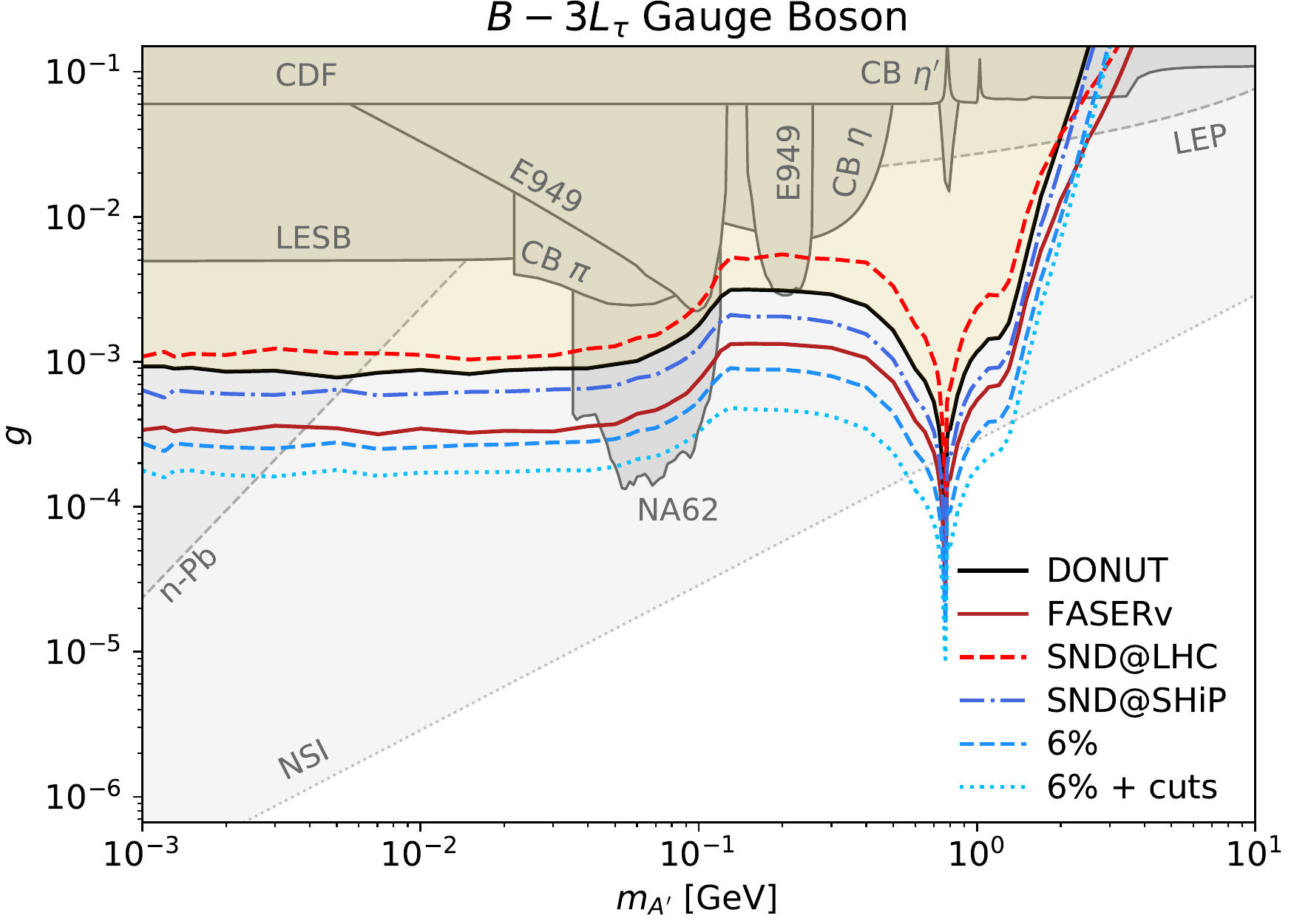}
\caption{
Regions of parameter space in the $B\!-\!L$ (upper left), $B\!-\!L_e\!-\!2L_\tau$ (upper right), $B\!-\!L_\mu\!-\!2L_\tau$ (bottom left) and $B\!-\!3L_\tau$ (bottom right) scenarios that can be constrained by the measurement of the tau neutrino rate at DONuT (yellow shaded area with solid black line), FASER$\nu$ (solid dark red line), SND@LHC (dashed light red line) and SND@SHiP (dot-dashed blue line) assuming a $33\%$ systematic uncertainty on the tau neutrino flux in the SM. In addition, we also show the possible sensitivity of SND@SHiP if a $6\%$ systematic uncertainty can be achieved (dashed blue line) and if additional cuts on the vertex location are applied (dotted blue line). Existing constraints from direct searches are shown in dark gray, while indirect constraints from scattering experiments and precision measurements are shown in lighter gray. The region accommodating the $(g-2)_\mu$ anomaly is shown as a green band.
}
\label{fig:results}
\end{figure*}
%------------------------------------------------------------

%----------------------------------
\subsection{Sensitivity Estimate}
%----------------------------------

Before looking at the full parameter space, let us consider the $B\!-\!3L_\tau$ model with $m_V=10~\mev$ and $g=10^{-3}$ as a benchmark model. The expected number of tau neutrinos produced via the decay of the vector boson and interacting with the detector, as well as the average energy of these neutrinos, is shown in the right block of \autoref{tab:experiments}. We can note that the ratio of the $\nu_\tau$ event rate from vector boson decay to the SM $\nu_\tau$ event rate is largest for the FASER$\nu$ experiment. This is due to its small transverse size which has been chosen to be similar to the angular spread of pions with $\tev$ energy, $\theta \sim \Lambda_\text{QCD}/\tev \sim 0.2~\mrad$, which are the source of both dark photons in FASER and muon neutrinos in FASER$\nu$. 

The left panel in \autoref{fig:distributions} shows the rate of tau neutrino interactions per unit volume, normalized to the prediction at the beam axis, as a function of displacement from the beam axis. We show the distributions for tau neutrinos from $V$ and $D_s$ decay as thick and thin lines, respectively. As expected, the tau neutrino rate is largest at the beam axis, and drops when moving away from it. Additionally, we can see that neutrinos produced in light vector boson decay are much more collimated around the beam axis than tau neutrinos from $D_s$ decay. We indicate the detector's radial coverage by the gray arrows: while DONuT, FASER$\nu$ and SND@SHiP are centered around the beam axis, the SND@LHC detector is displaced. It will therefore probe only the larger displacement tail of the tau neutrino beam, with typically lower energy and, hence, lower interaction cross section. This explains why its event rate in \autoref{tab:experiments} is significantly lower than for FASER$\nu$, especially for neutrinos from $V$ decay. In the right panel in \autoref{fig:distributions}, we show the energy distribution of tau neutrinos produced in vector boson decay and interacting with the detector. Note again that the SND@LHC and FASER$\nu$ experiments probe different parts of the tau neutrino energy spectrum.

In \autoref{fig:results} we show the sensitivities of the tau neutrino experiments alongside the existing constraints discussed in \autoref{sec:constraints}. The $B\!-\!L$ and $B\!-\!L_e\!-\!2L_\tau$ are strongly constrained by direct searches for both visible and invisible final decays of the vector boson, excluding couplings $g \gtrsim 3\cdot 10^{-4}$ over the whole considered mass range. In contrast, the direct constraints on the $B\!-\!L_\mu\!-\!2L_\tau$ and $B\!-\!3L_\tau$ models are much weaker, due to both the absence of $V$ production in electron experiments and the absence of the decay $V \to ee$. The leading bounds for these models come from indirect searches, for example from neutrino scattering or precision measurements. The strongest of these bounds is due to NSI constraints, which have been obtained by a global fit to neutrino oscillation data. As mentioned before, the indirect constraints are somewhat model dependent and could be relaxed in the presence of additional new physics. 

For each of the considered tau neutrino experiments, we require the predicted number of events from vector boson decay to be larger than twice the standard deviation of the SM prediction. The resulting event thresholds, $N_{2\sigma}$, are shown in the last column of \autoref{tab:experiments}. The recasted bounds for the DONuT experiment are shown as shaded yellow regions enclosed by solid black lines. We can identify an enhanced sensitivity at low masses $m_V \lesssim m_\pi$, where the vector boson can be abundantly produced in pion decay $\pi^0 \to V \gamma$, and at $m_V \approx m_\omega$, where its production is enhanced due to resonant mixing with the $\omega$ meson. At larger masses, $m_V\gtrsim 1~\gev$, the production cross section quickly drops. The DONuT bound is the strongest direct constraint for a large mass range in the $B\!-\!L_\mu\!-\!2L_\tau$ and $B\!-\!3L_\tau$ models, and the strongest constraint at $m_V\approx m_\omega$ for all models. 

The projected sensitivities of the FASER$\nu$, SND@LHC and SND@SHiP detectors are shown as solid dark red, dashed light red and dot-dashed blue lines, respectively. The FASER$\nu$ detector can extend the reach to roughly three times smaller couplings compared to DONuT. It benefits from a strongly collimated beam of neutrinos from vector boson decays, which is directly pointed at the detector. In contrast, the SND@LHC detector has a significantly weaker reach due to its offset from the beam axis which causes the peak of this neutrino beam to miss the detector, reducing its event rate. This effect is reduced at higher masses, $m>1~\gev$, where the two sensitivity curves come closer. Although the SND@SHiP proposal benefits from a much larger neutrino flux, its sensitivity is limited by systematic uncertainties, resulting in roughly the same reach as FASER$\nu$.

All considered experiments are limited by the assumed $33\%$ systematic uncertainties of the SM tau neutrino flux and an increased event rate will not lead to a significantly improved reach. Therefore, a better reach can  be obtained only when these flux uncertainties are reduced, for example, through a direct measurement of the tau neutrino production rate. In the case of SND@SHiP, this will be achieved by the recently approved DsTau~\cite{Aoki:2017spj, Aoki:2019jry} experiment at CERN's SPS. It will use an emulsion detector to measure the production rate of tau neutrinos in $D_s$ meson decay directly, and is expected to reduce the flux uncertainty to below $10\%$. To illustrate the impact of this measurement, we also show the reach of the SND@SHiP experiment with a $6\%$ systematic uncertainty as a dashed blue line in \autoref{fig:results}.

Finally, we note that differences in kinematic distributions, in particular the radial distribution around the beam collision axis, can be used to further enhance the sensitivity. While the tau neutrino event rates at DONuT, FASER$\nu$ and SND@LHC are generally low, the SND@SHiP detector will collect a large number of events and therefore be able to perform a shape analysis. We illustrate this by applying a cut on the event location and consider only neutrino interactions within the inner $20\cm \times 20\cm$ region of the detector. This cut increases the ratio of tau neutrino events from $V$ decay to those from $D_s$ decay by roughly a factor two. The resulting reach is shown as a dotted blue line in \autoref{fig:results}.

%%%%%%%%%%%%%%%%%%%%%%%%%%%%%%%%%%%%%
\section{Conclusion and Outlook}
\label{sec:summary}
%%%%%%%%%%%%%%%%%%%%%%%%%%%%%%%%%%%%%

In recent years, an extensive program has emerged to search for light and weakly interacting particles with masses in the $\mev - \gev$ range~\cite{Battaglieri:2017aum, Beacham:2019nyx}. Among their many motivations, such particles could help to explain the observed dark matter relic density and resolve anomalies in low-energy experiments~\cite{Bennett:2006fi, Pohl:2010zza,  Krasznahorkay:2015iga}. Searches from beam dump, fixed target, and collider experiments as well as neutrino scattering and precision measurements have been used to constrain these models, and a series of future searches and experiments will continue to search for signs of new physics associated with these models.

In this study, we have investigated the possibility of using the tau neutrino flux measurement to constrain models of light and weakly interacting particles. We have considered four models of light vector bosons associated with the anomaly-free $U(1)$ gauge groups of the $B\!-\!L$, $B\!-\!L_\mu\!-\!2L_\tau$, $B\!-\!L_e\!-\!2L_\tau$ and $B\!-\!3L_\tau$ numbers. These vector bosons can be produced in large numbers at high-energy experiments, for example through light meson decays such as $\pi^0 \to V \gamma$, and decay with an $\mathcal{O}(1)$ branching fraction into tau neutrinos. For comparison, in the SM only roughly one in $10^5$ high-energy hadron collisions leads to the production of a tau neutrino, meaning that even rare BSM processes could lead to sizable contributions to the tau neutrino flux. 

While neutrino interaction rates are naturally small, the identification of tau neutrinos further requires a detector with sufficient spatial resolution to identify the tau lepton in the final state. Tau neutrino experiments typically overcome this problem using emulsion detectors, which can achieve a sub-$\micm$ spatial resolution. In this study, we have considered four tau neutrino experiments: the DONuT experiment, which detected a total of 9 tau neutrino events, as well as the future FASER$\nu$, SND@LHC and SND@SHiP detectors and studied their sensitivity. We have found that DONuT imposes the strongest direct constraints in parts of the parameter space of the $B\!-\!L_e\!-\!2L_\tau$ and $B\!-\!3L_\tau$ models. In particular, for masses around $m_V\approx m_\omega$ the DONuT bounds exceed the indirect constraints arising from NSI measurements. The considered future tau neutrino experiments will further extend the sensitivity toward smaller couplings.

Finally, let us note once more that the proposed searches rely on an accurate understanding of the SM tau neutrino flux, which currently still has large uncertainties. This therefore motivates the further study of tau neutrino production through direct measurements~\cite{Aoki:2019jry}, precision QCD calculations~\cite{Bai:2020ukz} and improved simulators to understand and improve the flux uncertainties. 

%++++++++++++++++++++++++++++++++++++++++++++++++++++++++
%Acknowledgments
%++++++++++++++++++++++++++++++++++++++++++++++++++++++++

\acknowledgments
We thank Akitaka Ariga, Tomoko Ariga, Jonathan Feng, Julian Heeck, Ahmed Ismail, and Sebastian Trojanowski for useful discussions and comments on the manuscript. We are also grateful to the authors and maintainers of many open-source software packages, including
\texttt{CRMC}~\cite{CRMC},
\texttt{DarkCast}~\cite{Ilten:2018crw},
\texttt{EPOS}~\cite{Pierog:2013ria}
\texttt{Jupyter} notebooks~\cite{soton403913}, 
\texttt{Matplotlib}~\cite{Hunter:2007}, 
\texttt{NumPy}~\cite{numpy}, 
\texttt{pylhe}~\cite{lukas_2018_1217032}, 
\texttt{Pythia~8}~\cite{Sjostrand:2014zea},
\texttt{scikit-hep}~\cite{Rodrigues:2019nct}, 
and \texttt{Sibyll}~\cite{Fedynitch:2018cbl}.
FK is supported by the Department of Energy under Grant No. DE-AC02-76SF00515.

%++++++++++++++++++++++++++++++++++++++++++++++++++++++++
% References
%++++++++++++++++++++++++++++++++++++++++++++++++++++++++

\bibliography{references}

\providecommand{\href}[2]{#2}\begingroup\raggedright\begin{thebibliography}{10}

\bibitem{Nakamura:2006xs}
T.~Nakamura {\em et al.}, ``{The OPERA film: New nuclear emulsion for
  large-scale, high-precision experiments},''
  \href{http://dx.doi.org/10.1016/j.nima.2005.08.109}{{\em Nucl. Instrum. Meth.
  A} {\bf 556} (2006)  80--86}.

\bibitem{Agafonova:2018auq}
{\bf OPERA} Collaboration, N.~Agafonova {\em et al.}, ``{Final Results of the
  OPERA Experiment on $\nu_\tau$ Appearance in the CNGS Neutrino Beam},''
  \href{http://dx.doi.org/10.1103/PhysRevLett.121.139901,
  10.1103/PhysRevLett.120.211801}{{\em Phys. Rev. Lett.} {\bf 120} (2018)
  no.~21, 211801}, \href{http://arxiv.org/abs/1804.04912}{{\tt arXiv:1804.04912
  [hep-ex]}}.
[Erratum: Phys. Rev. Lett.121,no.13,139901(2018)].
%%CITATION = ARXIV:1804.04912;%%.

\bibitem{Kodama:2007aa}
{\bf DONuT} Collaboration, K.~Kodama {\em et al.}, ``{Final tau-neutrino
  results from the DONuT experiment},''
  \href{http://dx.doi.org/10.1103/PhysRevD.78.052002}{{\em Phys.\ Rev.\ D} {\bf
  78} (2008)  052002}, \href{http://arxiv.org/abs/0711.0728}{{\tt
  arXiv:0711.0728 [hep-ex]}}.

\bibitem{Anelli:2015pba}
{\bf SHiP} Collaboration, M.~Anelli {\em et al.}, ``{A facility to Search for
  Hidden Particles (SHiP) at the CERN SPS},''
  \href{http://arxiv.org/abs/1504.04956}{{\tt arXiv:1504.04956
  [physics.ins-det]}}.

\bibitem{Ahdida:2654870}
{\bf SHiP Collaboration} Collaboration, C.~Ahdida {\em et al.}, ``{SHiP
  Experiment - Progress Report},'' Tech. Rep. CERN-SPSC-2019-010. SPSC-SR-248,
  CERN, Geneva, Jan, 2019.
\newblock \url{https://cds.cern.ch/record/2654870}.

\bibitem{Ahdida:2704147}
{\bf SHiP Collaboration} Collaboration, C.~Ahdida {\em et al.}, ``{SHiP
  Experiment - Comprehensive Design Study report},'' Tech. Rep.
  CERN-SPSC-2019-049. SPSC-SR-263, CERN, Geneva, Dec, 2019.
\newblock \url{https://cds.cern.ch/record/2704147}.

\bibitem{Abreu:2019yak}
{\bf FASER} Collaboration, H.~Abreu {\em et al.}, ``{Detecting and Studying
  High-Energy Collider Neutrinos with FASER at the LHC},''
\href{http://arxiv.org/abs/1908.02310}{{\tt arXiv:1908.02310 [hep-ex]}}.
%%CITATION = ARXIV:1908.02310;%%.

\bibitem{Abreu:2020ddv}
{\bf FASER} Collaboration, H.~Abreu {\em et al.}, ``{Technical Proposal:
  FASERnu},''
\href{http://arxiv.org/abs/2001.03073}{{\tt arXiv:2001.03073
  [physics.ins-det]}}.
%%CITATION = ARXIV:2001.03073;%%.

\bibitem{Ahdida:2020evc}
{\bf SHiP} Collaboration, C.~Ahdida {\em et al.}, ``{SND@LHC},''
  \href{http://arxiv.org/abs/2002.08722}{{\tt arXiv:2002.08722
  [physics.ins-det]}}.

\bibitem{Araki:2012ip}
T.~Araki, J.~Heeck, and J.~Kubo, ``{Vanishing Minors in the Neutrino Mass
  Matrix from Abelian Gauge Symmetries},''
  \href{http://dx.doi.org/10.1007/JHEP07(2012)083}{{\em JHEP} {\bf 07} (2012)
  083}, \href{http://arxiv.org/abs/1203.4951}{{\tt arXiv:1203.4951 [hep-ph]}}.

\bibitem{Asai:2019ciz}
K.~Asai, ``{Predictions for the neutrino parameters in the minimal model
  extended by linear combination of U(1)$_{L_e-L_\mu}$, U(1)$_{L_\mu-L_\tau}$
  and U(1)$_{B-L}$ gauge symmetries},''
  \href{http://dx.doi.org/10.1140/epjc/s10052-020-7622-6}{{\em Eur. Phys. J.}
  {\bf C80} (2020) no.~2, 76},
\href{http://arxiv.org/abs/1907.04042}{{\tt arXiv:1907.04042 [hep-ph]}}.
%%CITATION = ARXIV:1907.04042;%%.

\bibitem{Bauer:2018onh}
M.~Bauer, P.~Foldenauer, and J.~Jaeckel, ``{Hunting All the Hidden Photons},''
  \href{http://dx.doi.org/10.1007/JHEP07(2018)094}{{\em JHEP} {\bf 07} (2018)
  094},
\href{http://arxiv.org/abs/1803.05466}{{\tt arXiv:1803.05466 [hep-ph]}}.
%%CITATION = ARXIV:1803.05466;%%.

\bibitem{Bahraminasr:2020ssz}
M.~Bahraminasr, P.~Bakhti, and M.~Rajaee, ``{Sensitivities to secret neutrino
  interaction at FASER$\nu$},''
\href{http://arxiv.org/abs/2003.09985}{{\tt arXiv:2003.09985 [hep-ph]}}.
%%CITATION = ARXIV:2003.09985;%%.

\bibitem{Ilten:2018crw}
P.~Ilten, Y.~Soreq, M.~Williams, and W.~Xue, ``{Serendipity in dark photon
  searches},'' \href{http://dx.doi.org/10.1007/JHEP06(2018)004}{{\em JHEP} {\bf
  06} (2018)  004}, \href{http://arxiv.org/abs/1801.04847}{{\tt
  arXiv:1801.04847 [hep-ph]}}.

\bibitem{Lees:2014xha}
{\bf BaBar} Collaboration, J.~P. Lees {\em et al.}, ``{Search for a Dark Photon
  in $e^+e^-$ Collisions at BaBar},''
  \href{http://dx.doi.org/10.1103/PhysRevLett.113.201801}{{\em Phys. Rev.
  Lett.} {\bf 113} (2014) no.~20, 201801},
\href{http://arxiv.org/abs/1406.2980}{{\tt arXiv:1406.2980 [hep-ex]}}.
%%CITATION = ARXIV:1406.2980;%%.

\bibitem{Aaij:2019bvg}
{\bf LHCb} Collaboration, R.~Aaij {\em et al.}, ``{Search for
  $A'\!\to\!\mu^+\mu^-$ decays},''
\href{http://arxiv.org/abs/1910.06926}{{\tt arXiv:1910.06926 [hep-ex]}}.
%%CITATION = ARXIV:1910.06926;%%.

\bibitem{Sirunyan:2018nnz}
{\bf CMS} Collaboration, A.~M. Sirunyan {\em et al.}, ``{Search for an
  $L_{\mu}-L_{\tau}$ gauge boson using Z$\to4\mu$ events in proton-proton
  collisions at $\sqrt{s} =$ 13 TeV},''
  \href{http://dx.doi.org/10.1016/j.physletb.2019.01.072}{{\em Phys. Lett. B}
  {\bf 792} (2019)  345--368}, \href{http://arxiv.org/abs/1808.03684}{{\tt
  arXiv:1808.03684 [hep-ex]}}.

\bibitem{Chun:2018ibr}
E.~J. Chun, A.~Das, J.~Kim, and J.~Kim, ``{Searching for flavored gauge
  bosons},'' \href{http://dx.doi.org/10.1007/JHEP02(2019)093}{{\em JHEP} {\bf
  02} (2019)  093}, \href{http://arxiv.org/abs/1811.04320}{{\tt
  arXiv:1811.04320 [hep-ph]}}. [Erratum: JHEP 07, 024 (2019)].

\bibitem{Blumlein:1990ay}
J.~Blumlein {\em et al.}, ``{Limits on neutral light scalar and pseudoscalar
  particles in a proton beam dump experiment},''
\href{http://dx.doi.org/10.1007/BF01548556}{{\em Z. Phys.} {\bf C51} (1991)
  341--350}.
%%CITATION = ZEPYA,C51,341;%%.

\bibitem{Blumlein:1991xh}
J.~Blumlein {\em et al.}, ``{Limits on the mass of light (pseudo)scalar
  particles from Bethe-Heitler e+ e- and mu+ mu- pair production in a proton -
  iron beam dump experiment},''
\href{http://dx.doi.org/10.1142/S0217751X9200171X}{{\em Int. J. Mod. Phys.}
  {\bf A7} (1992)  3835--3850}.
%%CITATION = IMPAE,A7,3835;%%.

\bibitem{Davier:1989wz}
M.~Davier and H.~Nguyen~Ngoc, ``{An Unambiguous Search for a Light Higgs
  Boson},''
\href{http://dx.doi.org/10.1016/0370-2693(89)90174-3}{{\em Phys. Lett.} {\bf
  B229} (1989)  150--155}.
%%CITATION = PHLTA,B229,150;%%.

\bibitem{Lees:2017lec}
{\bf BaBar} Collaboration, J.~P. Lees {\em et al.}, ``{Search for Invisible
  Decays of a Dark Photon Produced in ${e}^{+}{e}^{-}$ Collisions at BaBar},''
  \href{http://dx.doi.org/10.1103/PhysRevLett.119.131804}{{\em Phys. Rev.
  Lett.} {\bf 119} (2017) no.~13, 131804},
\href{http://arxiv.org/abs/1702.03327}{{\tt arXiv:1702.03327 [hep-ex]}}.
%%CITATION = ARXIV:1702.03327;%%.

\bibitem{NA64:2019imj}
D.~Banerjee {\em et al.}, ``{Dark matter search in missing energy events with
  NA64},'' \href{http://dx.doi.org/10.1103/PhysRevLett.123.121801}{{\em Phys.
  Rev. Lett.} {\bf 123} (2019) no.~12, 121801},
\href{http://arxiv.org/abs/1906.00176}{{\tt arXiv:1906.00176 [hep-ex]}}.
%%CITATION = ARXIV:1906.00176;%%.

\bibitem{CortinaGil:2019nuo}
{\bf NA62} Collaboration, E.~Cortina~Gil {\em et al.}, ``{Search for production
  of an invisible dark photon in $\pi^0$ decays},''
  \href{http://dx.doi.org/10.1007/JHEP05(2019)182}{{\em JHEP} {\bf 05} (2019)
  182}, \href{http://arxiv.org/abs/1903.08767}{{\tt arXiv:1903.08767
  [hep-ex]}}.

\bibitem{Atiya:1992sm}
M.~Atiya {\em et al.}, ``{Search for the decay $\pi^0 \to \gamma X$},''
  \href{http://dx.doi.org/10.1103/PhysRevLett.69.733}{{\em Phys. Rev. Lett.}
  {\bf 69} (1992)  733--736}.

\bibitem{Amsler:1994gt}
{\bf Crystal Barrel} Collaboration, C.~Amsler {\em et al.}, ``{Search for a new
  light gauge boson in decays of pi0 and eta},''
  \href{http://dx.doi.org/10.1016/0370-2693(94)91043-X}{{\em Phys. Lett. B}
  {\bf 333} (1994)  271--276}.

\bibitem{Artamonov:2009sz}
{\bf BNL-E949} Collaboration, A.~Artamonov {\em et al.}, ``{Study of the decay
  $K^+\to\pi^+\nu \bar\nu$ in the momentum region $140 < P_\pi < 199$ MeV/c},''
  \href{http://dx.doi.org/10.1103/PhysRevD.79.092004}{{\em Phys. Rev. D} {\bf
  79} (2009)  092004}, \href{http://arxiv.org/abs/0903.0030}{{\tt
  arXiv:0903.0030 [hep-ex]}}.

\bibitem{Pospelov:2008zw}
M.~Pospelov, ``{Secluded U(1) below the weak scale},''
  \href{http://dx.doi.org/10.1103/PhysRevD.80.095002}{{\em Phys. Rev. D} {\bf
  80} (2009)  095002}, \href{http://arxiv.org/abs/0811.1030}{{\tt
  arXiv:0811.1030 [hep-ph]}}.

\bibitem{Batell:2014yra}
B.~Batell, P.~deNiverville, D.~McKeen, M.~Pospelov, and A.~Ritz, ``{Leptophobic
  Dark Matter at Neutrino Factories},''
  \href{http://dx.doi.org/10.1103/PhysRevD.90.115014}{{\em Phys. Rev. D} {\bf
  90} (2014) no.~11, 115014}, \href{http://arxiv.org/abs/1405.7049}{{\tt
  arXiv:1405.7049 [hep-ph]}}.

\bibitem{Aaltonen:2012jb}
{\bf CDF} Collaboration, T.~Aaltonen {\em et al.}, ``{A Search for dark matter
  in events with one jet and missing transverse energy in $p\bar{p}$ collisions
  at $\sqrt{s} = 1.96$ TeV},''
  \href{http://dx.doi.org/10.1103/PhysRevLett.108.211804}{{\em Phys. Rev.
  Lett.} {\bf 108} (2012)  211804}, \href{http://arxiv.org/abs/1203.0742}{{\tt
  arXiv:1203.0742 [hep-ex]}}.

\bibitem{Shoemaker:2011vi}
I.~M. Shoemaker and L.~Vecchi, ``{Unitarity and Monojet Bounds on Models for
  DAMA, CoGeNT, and CRESST-II},''
  \href{http://dx.doi.org/10.1103/PhysRevD.86.015023}{{\em Phys. Rev. D} {\bf
  86} (2012)  015023}, \href{http://arxiv.org/abs/1112.5457}{{\tt
  arXiv:1112.5457 [hep-ph]}}.

\bibitem{Mishra:1991bv}
{\bf CCFR} Collaboration, S.~Mishra {\em et al.}, ``{Neutrino tridents and W Z
  interference},'' \href{http://dx.doi.org/10.1103/PhysRevLett.66.3117}{{\em
  Phys. Rev. Lett.} {\bf 66} (1991)  3117--3120}.

\bibitem{Altmannshofer:2014pba}
W.~Altmannshofer, S.~Gori, M.~Pospelov, and I.~Yavin, ``{Neutrino Trident
  Production: A Powerful Probe of New Physics with Neutrino Beams},''
  \href{http://dx.doi.org/10.1103/PhysRevLett.113.091801}{{\em Phys. Rev.
  Lett.} {\bf 113} (2014)  091801},
\href{http://arxiv.org/abs/1406.2332}{{\tt arXiv:1406.2332 [hep-ph]}}.
%%CITATION = ARXIV:1406.2332;%%.

\bibitem{Vilain:1993kd}
{\bf CHARM-II} Collaboration, P.~Vilain {\em et al.}, ``{Measurement of
  differential cross-sections for muon-neutrino electron scattering},''
  \href{http://dx.doi.org/10.1016/0370-2693(93)90408-A}{{\em Phys. Lett. B}
  {\bf 302} (1993)  351--355}.

\bibitem{Bilmis:2015lja}
S.~Bilmis, I.~Turan, T.~M. Aliev, M.~Deniz, L.~Singh, and H.~T. Wong,
  ``{Constraints on Dark Photon from Neutrino-Electron Scattering
  Experiments},'' \href{http://dx.doi.org/10.1103/PhysRevD.92.033009}{{\em
  Phys. Rev.} {\bf D92} (2015) no.~3, 033009},
\href{http://arxiv.org/abs/1502.07763}{{\tt arXiv:1502.07763 [hep-ph]}}.
%%CITATION = ARXIV:1502.07763;%%.

\bibitem{Akimov:2017ade}
{\bf COHERENT} Collaboration, D.~Akimov {\em et al.}, ``{Observation of
  Coherent Elastic Neutrino-Nucleus Scattering},''
  \href{http://dx.doi.org/10.1126/science.aao0990}{{\em Science} {\bf 357}
  (2017) no.~6356, 1123--1126}, \href{http://arxiv.org/abs/1708.01294}{{\tt
  arXiv:1708.01294 [nucl-ex]}}.

\bibitem{Kosmas:2017tsq}
D.~Papoulias and T.~Kosmas, ``{COHERENT constraints to conventional and exotic
  neutrino physics},'' \href{http://dx.doi.org/10.1103/PhysRevD.97.033003}{{\em
  Phys. Rev. D} {\bf 97} (2018) no.~3, 033003},
  \href{http://arxiv.org/abs/1711.09773}{{\tt arXiv:1711.09773 [hep-ph]}}.

\bibitem{Bennett:2006fi}
{\bf Muon g-2} Collaboration, G.~Bennett {\em et al.}, ``{Final Report of the
  Muon E821 Anomalous Magnetic Moment Measurement at BNL},''
  \href{http://dx.doi.org/10.1103/PhysRevD.73.072003}{{\em Phys. Rev. D} {\bf
  73} (2006)  072003}, \href{http://arxiv.org/abs/hep-ex/0602035}{{\tt
  arXiv:hep-ex/0602035}}.

\bibitem{Tanabashi:2018oca}
{\bf Particle Data Group} Collaboration, M.~Tanabashi {\em et al.}, ``{Review
  of Particle Physics},''
  \href{http://dx.doi.org/10.1103/PhysRevD.98.030001}{{\em Phys. Rev. D} {\bf
  98} (2018) no.~3, 030001}.

\bibitem{Ma:1998dp}
E.~Ma and D.~Roy, ``{Phenomenology of the $B$ - 3L($\tau$) gauge boson},''
  \href{http://dx.doi.org/10.1103/PhysRevD.58.095005}{{\em Phys. Rev. D} {\bf
  58} (1998)  095005}, \href{http://arxiv.org/abs/hep-ph/9806210}{{\tt
  arXiv:hep-ph/9806210}}.

\bibitem{Barbieri:1975xy}
R.~Barbieri and T.~E.~O. Ericson, ``{Evidence Against the Existence of a Low
  Mass Scalar Boson from Neutron-Nucleus Scattering},''
  \href{http://dx.doi.org/10.1016/0370-2693(75)90073-8}{{\em Phys. Lett. B}
  {\bf 57} (1975)  270--272}.

\bibitem{Barger:2010aj}
V.~Barger, C.-W. Chiang, W.-Y. Keung, and D.~Marfatia, ``{Proton size
  anomaly},'' \href{http://dx.doi.org/10.1103/PhysRevLett.106.153001}{{\em
  Phys. Rev. Lett.} {\bf 106} (2011)  153001},
  \href{http://arxiv.org/abs/1011.3519}{{\tt arXiv:1011.3519 [hep-ph]}}.

\bibitem{Esteban:2018ppq}
I.~Esteban, M.~Gonzalez-Garcia, M.~Maltoni, I.~Martinez-Soler, and J.~Salvado,
  ``{Updated Constraints on Non-Standard Interactions from Global Analysis of
  Oscillation Data},'' \href{http://dx.doi.org/10.1007/JHEP08(2018)180}{{\em
  JHEP} {\bf 08} (2018)  180}, \href{http://arxiv.org/abs/1805.04530}{{\tt
  arXiv:1805.04530 [hep-ph]}}.

\bibitem{Heeck:2018nzc}
J.~Heeck, M.~Lindner, W.~Rodejohann, and S.~Vogl, ``{Non-Standard Neutrino
  Interactions and Neutral Gauge Bosons},''
  \href{http://dx.doi.org/10.21468/SciPostPhys.6.3.038}{{\em SciPost Phys.}
  {\bf 6} (2019) no.~3, 038}, \href{http://arxiv.org/abs/1812.04067}{{\tt
  arXiv:1812.04067 [hep-ph]}}.

\bibitem{Foldenauer:2019dai}
P.~Foldenauer, ``{Dark Sectors from the Hidden Photon Perspective},'' in {\em
  {7th Large Hadron Collider Physics Conference (LHCP 2019) Puebla, Puebla,
  Mexico, May 20-25, 2019}}.
\newblock 2019.
\newblock
\href{http://arxiv.org/abs/1907.10630}{{\tt arXiv:1907.10630 [hep-ph]}}.
\newblock
%%CITATION = ARXIV:1907.10630;%%.

\bibitem{Kodama:2000mp}
{\bf DONUT} Collaboration, K.~Kodama {\em et al.}, ``{Observation of tau
  neutrino interactions},''
  \href{http://dx.doi.org/10.1016/S0370-2693(01)00307-0}{{\em Phys. Lett. B}
  {\bf 504} (2001)  218--224}, \href{http://arxiv.org/abs/hep-ex/0012035}{{\tt
  arXiv:hep-ex/0012035}}.

\bibitem{Feng:2017vli}
J.~L. Feng, I.~Galon, F.~Kling, and S.~Trojanowski, ``{Dark Higgs bosons at the
  ForwArd Search ExpeRiment},''
  \href{http://dx.doi.org/10.1103/PhysRevD.97.055034}{{\em Phys. Rev.} {\bf
  D97} (2018) no.~5, 055034},
\href{http://arxiv.org/abs/1710.09387}{{\tt arXiv:1710.09387 [hep-ph]}}.
%%CITATION = ARXIV:1710.09387;%%.

\bibitem{Feng:2017uoz}
J.~L. Feng, I.~Galon, F.~Kling, and S.~Trojanowski, ``{ForwArd Search
  ExpeRiment at the LHC},''
  \href{http://dx.doi.org/10.1103/PhysRevD.97.035001}{{\em Phys. Rev. D} {\bf
  97} (2018) no.~3, 035001}, \href{http://arxiv.org/abs/1708.09389}{{\tt
  arXiv:1708.09389 [hep-ph]}}.

\bibitem{Feng:2018noy}
J.~L. Feng, I.~Galon, F.~Kling, and S.~Trojanowski, ``{Axionlike particles at
  FASER: The LHC as a photon beam dump},''
  \href{http://dx.doi.org/10.1103/PhysRevD.98.055021}{{\em Phys. Rev.} {\bf
  D98} (2018) no.~5, 055021},
\href{http://arxiv.org/abs/1806.02348}{{\tt arXiv:1806.02348 [hep-ph]}}.
%%CITATION = ARXIV:1806.02348;%%.

\bibitem{Kling:2018wct}
F.~Kling and S.~Trojanowski, ``{Heavy Neutral Leptons at FASER},''
  \href{http://dx.doi.org/10.1103/PhysRevD.97.095016}{{\em Phys. Rev.} {\bf
  D97} (2018) no.~9, 095016},
\href{http://arxiv.org/abs/1801.08947}{{\tt arXiv:1801.08947 [hep-ph]}}.
%%CITATION = ARXIV:1801.08947;%%.

\bibitem{Berlin:2018jbm}
A.~Berlin and F.~Kling, ``{Inelastic Dark Matter at the LHC Lifetime Frontier:
  ATLAS, CMS, LHCb, CODEX-b, FASER, and MATHUSLA},''
  \href{http://dx.doi.org/10.1103/PhysRevD.99.015021}{{\em Phys. Rev.} {\bf
  D99} (2019) no.~1, 015021},
\href{http://arxiv.org/abs/1810.01879}{{\tt arXiv:1810.01879 [hep-ph]}}.
%%CITATION = ARXIV:1810.01879;%%.

\bibitem{Ariga:2018zuc}
{\bf FASER} Collaboration, A.~Ariga {\em et al.}, ``{Letter of Intent for
  FASER: ForwArd Search ExpeRiment at the LHC},''
  \href{http://arxiv.org/abs/1811.10243}{{\tt arXiv:1811.10243
  [physics.ins-det]}}. \url{https://cds.cern.ch/record/2642351}.
Submitted to the CERN LHCC on 18 July 2018.
%%CITATION = ARXIV:1811.10243;%%.

\bibitem{Ariga:2018uku}
{\bf FASER} Collaboration, A.~Ariga {\em et al.}, ``{FASER’s physics reach
  for long-lived particles},''
  \href{http://dx.doi.org/10.1103/PhysRevD.99.095011}{{\em Phys. Rev.} {\bf
  D99} (2019) no.~9, 095011},
\href{http://arxiv.org/abs/1811.12522}{{\tt arXiv:1811.12522 [hep-ph]}}.
%%CITATION = ARXIV:1811.12522;%%.

\bibitem{Ariga:2018pin}
{\bf FASER} Collaboration, A.~Ariga {\em et al.}, ``{Technical Proposal for
  FASER: ForwArd Search ExpeRiment at the LHC},''
  \href{http://arxiv.org/abs/1812.09139}{{\tt arXiv:1812.09139
  [physics.ins-det]}}. \url{http://cds.cern.ch/record/2651328}.
Submitted to the CERN LHCC on 7 November 2018.
%%CITATION = ARXIV:1812.09139;%%.

\bibitem{Ariga:2019ufm}
{\bf FASER} Collaboration, A.~Ariga {\em et al.}, ``{FASER: ForwArd Search
  ExpeRiment at the LHC},'' \href{http://arxiv.org/abs/1901.04468}{{\tt
  arXiv:1901.04468 [hep-ex]}}.

\bibitem{Alekhin:2015byh}
S.~Alekhin {\em et al.}, ``{A facility to Search for Hidden Particles at the
  CERN SPS: the SHiP physics case},''
  \href{http://dx.doi.org/10.1088/0034-4885/79/12/124201}{{\em Rept. Prog.
  Phys.} {\bf 79} (2016) no.~12, 124201},
  \href{http://arxiv.org/abs/1504.04855}{{\tt arXiv:1504.04855 [hep-ph]}}.

\bibitem{Engel:2015dxa}
F.~Riehn, R.~Engel, A.~Fedynitch, T.~K. Gaisser, and T.~Stanev, ``{Charm
  production in SIBYLL},''
  \href{http://dx.doi.org/10.1051/epjconf/20159912001}{{\em EPJ Web Conf.} {\bf
  99} (2015)  12001}, \href{http://arxiv.org/abs/1502.06353}{{\tt
  arXiv:1502.06353 [hep-ph]}}.

\bibitem{Fedynitch:2018cbl}
A.~Fedynitch, F.~Riehn, R.~Engel, T.~K. Gaisser, and T.~Stanev, ``{Hadronic
  interaction model sibyll 2.3c and inclusive lepton fluxes},''
  \href{http://dx.doi.org/10.1103/PhysRevD.100.103018}{{\em Phys. Rev. D} {\bf
  100} (2019) no.~10, 103018}, \href{http://arxiv.org/abs/1806.04140}{{\tt
  arXiv:1806.04140 [hep-ph]}}.

\bibitem{Sjostrand:2014zea}
T.~Sjöstrand, S.~Ask, J.~R. Christiansen, R.~Corke, N.~Desai, P.~Ilten,
  S.~Mrenna, S.~Prestel, C.~O. Rasmussen, and P.~Z. Skands, ``{An Introduction
  to PYTHIA 8.2},'' \href{http://dx.doi.org/10.1016/j.cpc.2015.01.024}{{\em
  Comput. Phys. Commun.} {\bf 191} (2015)  159--177},
  \href{http://arxiv.org/abs/1410.3012}{{\tt arXiv:1410.3012 [hep-ph]}}.

\bibitem{Skands:2014pea}
P.~Skands, S.~Carrazza, and J.~Rojo, ``{Tuning PYTHIA 8.1: the Monash 2013
  Tune},'' \href{http://dx.doi.org/10.1140/epjc/s10052-014-3024-y}{{\em Eur.
  Phys. J. C} {\bf 74} (2014) no.~8, 3024},
  \href{http://arxiv.org/abs/1404.5630}{{\tt arXiv:1404.5630 [hep-ph]}}.

\bibitem{ATL-PHYS-PUB-2012-003}
{\bf ATLAS} Collaboration, ``{Summary of ATLAS Pythia 8 tunes},'' Tech. Rep.
  ATL-PHYS-PUB-2012-003, CERN, Geneva, Aug, 2012.
\newblock \url{https://cds.cern.ch/record/1474107}.

\bibitem{Roesler:2000he}
S.~Roesler, R.~Engel, and J.~Ranft,
  \href{http://dx.doi.org/10.1007/978-3-642-18211-2\_166}{``{The Monte Carlo
  event generator DPMJET-III},''} in {\em {International Conference on Advanced
  Monte Carlo for Radiation Physics, Particle Transport Simulation and
  Applications (MC 2000)}}, pp.~1033--1038.
\newblock 12, 2000.
\newblock \href{http://arxiv.org/abs/hep-ph/0012252}{{\tt
  arXiv:hep-ph/0012252}}.

\bibitem{Bai:2018xum}
W.~Bai and M.~H. Reno, ``{Prompt neutrinos and intrinsic charm at SHiP},''
  \href{http://dx.doi.org/10.1007/JHEP02(2019)077}{{\em JHEP} {\bf 02} (2019)
  077}, \href{http://arxiv.org/abs/1807.02746}{{\tt arXiv:1807.02746
  [hep-ph]}}.

\bibitem{Bai:2020ukz}
W.~Bai, M.~Diwan, M.~V. Garzelli, Y.~S. Jeong, and M.~H. Reno, ``{Far-forward
  neutrinos at the Large Hadron Collider},''
  \href{http://arxiv.org/abs/2002.03012}{{\tt arXiv:2002.03012 [hep-ph]}}.

\bibitem{Aoki:2019jry}
{\bf DsTau} Collaboration, S.~Aoki {\em et al.}, ``{DsTau: Study of tau
  neutrino production with 400 GeV protons from the CERN-SPS},''
  \href{http://dx.doi.org/10.1007/JHEP01(2020)033}{{\em JHEP} {\bf 01} (2020)
  033}, \href{http://arxiv.org/abs/1906.03487}{{\tt arXiv:1906.03487
  [hep-ex]}}.

\bibitem{Aaij:2015bpa}
{\bf LHCb} Collaboration, R.~Aaij {\em et al.}, ``{Measurements of prompt charm
  production cross-sections in $pp$ collisions at $ \sqrt{s}=13 $ TeV},''
  \href{http://dx.doi.org/10.1007/JHEP03(2016)159}{{\em JHEP} {\bf 03} (2016)
  159}, \href{http://arxiv.org/abs/1510.01707}{{\tt arXiv:1510.01707
  [hep-ex]}}. [Erratum: JHEP 09, 013 (2016), Erratum: JHEP 05, 074 (2017)].

\bibitem{Pierog:2013ria}
T.~Pierog, I.~Karpenko, J.~Katzy, E.~Yatsenko, and K.~Werner, ``{EPOS LHC: Test
  of collective hadronization with data measured at the CERN Large Hadron
  Collider},'' \href{http://dx.doi.org/10.1103/PhysRevC.92.034906}{{\em Phys.
  Rev. C} {\bf 92} (2015) no.~3, 034906},
  \href{http://arxiv.org/abs/1306.0121}{{\tt arXiv:1306.0121 [hep-ph]}}.

\bibitem{CRMC}
C.~Baus, T.~Pierog, and R.~Ulrich, ``{Cosmic Ray Monte Carlo (CRMC)},''.
  \url{https://web.ikp.kit.edu/rulrich/crmc.html}.

\bibitem{Tulin:2014tya}
S.~Tulin, ``{New weakly-coupled forces hidden in low-energy QCD},''
  \href{http://dx.doi.org/10.1103/PhysRevD.89.114008}{{\em Phys. Rev. D} {\bf
  89} (2014) no.~11, 114008}, \href{http://arxiv.org/abs/1404.4370}{{\tt
  arXiv:1404.4370 [hep-ph]}}.

\bibitem{Sjostrand:2006za}
T.~Sjostrand, S.~Mrenna, and P.~Z. Skands, ``{PYTHIA 6.4 Physics and Manual},''
  \href{http://dx.doi.org/10.1088/1126-6708/2006/05/026}{{\em JHEP} {\bf 05}
  (2006)  026}, \href{http://arxiv.org/abs/hep-ph/0603175}{{\tt
  arXiv:hep-ph/0603175}}.

\bibitem{Aoki:2017spj}
S.~Aoki {\em et al.}, ``{Study of tau-neutrino production at the CERN SPS},''
  \href{http://arxiv.org/abs/1708.08700}{{\tt arXiv:1708.08700 [hep-ex]}}.

\bibitem{Battaglieri:2017aum}
M.~Battaglieri {\em et al.}, ``{US Cosmic Visions: New Ideas in Dark Matter
  2017: Community Report},'' in {\em {U.S. Cosmic Visions: New Ideas in Dark
  Matter}}.
\newblock 7, 2017.
\newblock \href{http://arxiv.org/abs/1707.04591}{{\tt arXiv:1707.04591
  [hep-ph]}}.

\bibitem{Beacham:2019nyx}
J.~Beacham {\em et al.}, ``{Physics Beyond Colliders at CERN: Beyond the
  Standard Model Working Group Report},''
  \href{http://dx.doi.org/10.1088/1361-6471/ab4cd2}{{\em J. Phys. G} {\bf 47}
  (2020) no.~1, 010501}, \href{http://arxiv.org/abs/1901.09966}{{\tt
  arXiv:1901.09966 [hep-ex]}}.

\bibitem{Pohl:2010zza}
R.~Pohl {\em et al.}, ``{The size of the proton},''
  \href{http://dx.doi.org/10.1038/nature09250}{{\em Nature} {\bf 466} (2010)
  213--216}.

\bibitem{Krasznahorkay:2015iga}
A.~J. Krasznahorkay {\em et al.}, ``{Observation of Anomalous Internal Pair
  Creation in Be8 : A Possible Indication of a Light, Neutral Boson},''
  \href{http://dx.doi.org/10.1103/PhysRevLett.116.042501}{{\em Phys. Rev.
  Lett.} {\bf 116} (2016) no.~4, 042501},
  \href{http://arxiv.org/abs/1504.01527}{{\tt arXiv:1504.01527 [nucl-ex]}}.

\bibitem{soton403913}
T.~Kluyver {\em et al.}, ``Jupyter notebooks - a publishing format for
  reproducible computational workflows,'' in {\em Positioning and Power in
  Academic Publishing: Players, Agents and Agendas}, pp.~87--90.
\newblock IOS Press, 2016.
\newblock \url{https://eprints.soton.ac.uk/403913/}.

\bibitem{Hunter:2007}
J.~D. Hunter, ``Matplotlib: A 2d graphics environment,''
  \href{http://dx.doi.org/10.1109/MCSE.2007.55}{{\em Computing in Science \&
  Engineering} {\bf 9} (2007) no.~3, 90--95}.

\bibitem{numpy}
T.~Oliphant, ``{NumPy}: A guide to {NumPy}.'' Usa: Trelgol publishing, 2006--.
\newblock \url{http://www.numpy.org/}.

\bibitem{lukas_2018_1217032}
L.~Heinrich, ``lukasheinrich/pylhe v0.0.4,'' Apr., 2018.
\newblock \url{https://doi.org/10.5281/zenodo.1217032}.

\bibitem{Rodrigues:2019nct}
E.~Rodrigues, ``{The Scikit-HEP Project},'' in {\em {23rd International
  Conference on Computing in High Energy and Nuclear Physics (CHEP 2018) Sofia,
  Bulgaria, July 9-13, 2018}}.
\newblock 2019.
\newblock
\href{http://arxiv.org/abs/1905.00002}{{\tt arXiv:1905.00002
  [physics.comp-ph]}}.
\newblock
%%CITATION = ARXIV:1905.00002;%%.

\end{thebibliography}\endgroup

\end{document}